\documentstyle[times,pramana,epsf,floats]{ias}
\newcommand{\cosb}{c_\beta}
\newcommand{\sinb}{s_\beta}
\newcommand{\gsim}{\buildrel>\over{_\sim}}
\begin{document}
\mark{{SUSY DM}{M.~Nojiri}}
\title{SUSY dark matter\footnote{a Japanese rescue operation {\ttfamily :-)}} \\
--{\it \small A Collider Physicist's Perspective}--}

\author{Mihoko M.~Nojiri}
\address{YITP, Kyoto University, Sakyo, Kyoto 606-8502, Japan}
\abstract{A short tour of supersymmetric dark matter 
and its connection to  collider physics.}
\maketitle

\section{Quest for the dark matter}

Recently the existence of dark matter has been confirmed through the CMB
measurements\cite{Bennett:2003bz}.
Because the baryon density of the universe 
$\Omega_b$ is only 1/6 of
the matter density of the universe  $\Omega_m
h^2=0.27\pm0.04$, the dark matter  should  be explained
by the physics beyond standard model(SM). 
There are numerous on-going and  planned dark matter
search experiments which aim to obtain direct evidence of  new
particles that constitute the dark matter in the universe.

On the other hand, high energy  colliders, such as Large Hadron Collider 
at CERN (LHC) or proposed Linear
$e^+e^-$ colliders (LC), would significantly extend our ability to explore
new physics at the TeV scale.  The two approaches--- collider experiments
and dark matter searches--- will open a new regime of the particle
physics in this century.

Supersymmetry (SUSY) is one of the favorable candidates for new physics at 
the TeV
scale. This is the only non-trivial extension of the Poincare algebra,
and the
it solves the hierarchy problem of the SM.  The minimal
supersymmetric extension of standard model (MSSM) is going to  be explored up
to a few TeV by LHC, and current and future dark matter searches also
have sensitivity to supersymmetric dark matter with the mass up to
the TeV scale.

In the first part of my talk, I  summarize   SUSY
dark matter searches.  Once a dark matter signal is observed, 
the data can be used to study  
the density profile of our galaxy, which is not fully
understood right now. In the latter half of my talk, I  discuss the
SUSY searches at colliders and more involved studies to measure the masses
and interactions of the sparticles.  Through the measurements, the thermal
relic density and  reaction of the SUSY dark matter would be
constrained severely. These will be solid bases to discuss
astrophysics and cosmology involving SUSY dark matter.

\section{From the Universe}

\subsection{Direct detections and local dark matter velocity distribution}
For cosmologists, dark matter is important because it has created the 
structure of the universe through its gravitational 
interaction.
Particle physicists see the same dark matter differently.
They view the dark matter as weakly interacting particles, 
and search for them through their interactions.

In the MSSM, the  lightest supersymmetric particle (LSP) is  
stable if R parity is conserved, and it is a candidate for 
the dark matter in the universe.  The LSP may be the lightest neutralino
$\tilde{\chi}^0_1$. The $\tilde{\chi}^0_1$  is a mixture of the 
superpartners of the gauge bosons ($\tilde{B}$ and $\tilde{W}$) and 
superpartners of the Higgs bosons ($\tilde{H}_1$ and $\tilde{H}_2$),
\begin{equation}
\tilde{\chi} (\equiv \tilde{\chi}^0_1)= 
N_{\tilde{B}}\tilde{B}+ N_{\tilde{W}}\tilde{W}
+N_{\tilde{H}_1}\tilde{H}_1+ N_{\tilde{H}_1}\tilde{H}_1. 
\end{equation}
The masses and mixings of the neutralinos $\tilde{\chi}^0_i$ ($i=1,..4$)
are determined by  $M_1$ (Bino mass)
$M_2$(Wino mass parameter) and Higgsino mass parameter $\mu$, and they 
also have  weak dependence on the ratio of the vacuum expectation 
value of the Higgs bosons $\tan\beta$. The mass matrix is given as
follows, 
\label{Sneumass}
\begin{equation}
{\cal M}_N =
\left( \begin{array}{cccc}
M_1             &0            &\! \! -m_Z \cosb\, s_W & m_Z \sinb\, s_W \\
0               &M_2          & \! \! m_Z \cosb\, c_W &-m_Z \sinb\, c_W \\
-m_Z \cosb\, s_W  & \! \! m_Z \cosb\, c_W &0            &-\mu           \\
 m_Z \sinb\, s_W  &\! \! -m_Z \sinb\, c_W &-\mu         &0     \end{array}
\right) ,
\end{equation}

The $\tilde{\chi}$ may be  directly searched for
through   $\tilde{\chi}N$ scattering 
mediated by a higgs  boson exchange. 
On the other hand, the  
superpartner of the graviton (gravitino $\tilde{\psi}_{3/2}$) is   the LSP
in the {\it gauge mediation model}.  The interaction of 
gravitino dark matter 
is too small  to obtain 
direct evidence for its existence.

The CDMS II is an exciting dark matter search experiment 
in the forthcoming few years. 
The CDMS II aims  to find a dark matter particle 
with the cross section 
$\sigma(N\tilde{\chi}\rightarrow N\tilde{\chi})>10^{-8}$pb 
( or $ 3\times 10^{-4}/({\rm kg\cdot keV \cdot day})$), while the 
current limit is around $10^{-6}$ pb. 
The Cryogenic detector  measures both phonon 
and ionization, therefore it actively discriminates 
the background caused by  electrons or photons. The remaining neutron 
background at the Soudan mine (the depth of 2090 mwe) is significantly 
lower than for
the previous CDMS experiment\cite{Abrams:2002nb} at a shallow site. 
As we have already heard in Roskowski's talk\cite{ros}, 
CDMS II cuts into the  significant region of the  
MSSM parameter space which is not excluded by current experimental 
constraints. In addition, there is a  hope that 
the dark matter signal rate is very close to the current limit, 
because of the claimed  ``evidence'' of the dark matter at DAMA
\cite{Bernabei:2000qi}. 
It is therefore tempting to think about the implication of the 
dark matter signal when  we have 
O(100) events at such high-tech detectors. 

Counting rate  of the dark matter depends on the velocity 
distribution. The standard assumption is
virialized dark matter. The velocity distribution is  
Gaussian and  the average velocity is zero.  Therefore, 
the average dark matter velocity is roughly equal to the earth
velocity $\sim 220$~km/s. Because of the earth motion around 
the sun, the  dark matter signal  modulates annually.  The modulation  is 
important to extract  the fraction of the  counting rate that  
comes from the dark matter scattering off the nuclei.

Currently we do not have any direct evidence  
of the dark matter velocity distribution in our galaxy. 
Dark matter in our galaxy may be co-rotating or counter-rotating around 
the center of the galaxy, and the dark matter flux significantly 
changes depending on that. After the discovery of the dark matter, 
the dark matter velocity distribution 
would be obtained experimentally  by studying  
the energy deposit in the detector\cite{Kamionkowski:1997xg}.
 
An even more exciting possibility was pointed out by Sikivie et al
\cite{Sikivie:1996nn}. 
They assume a non-virialized dark halo in which the collisionless 
dark matter particles falling into the galaxy oscillate in and out many
times. The expected phase of the annual modulation of a 
dark matter signal in direct detection experiments 
is opposite to the one usually expected\cite{Gelmini:2000dm}.

\subsection{Indirect detections and the dark matter profile in the galaxy}

Exotic cosmic rays such as $\gamma$, $\bar{p}$ and  $e^+$
are produced by the neutralino dark matter pair annihilation in the galaxy. 
They may be observed by space or balloon-borne experiments, and 
are called  {\it indirect detections} of the dark matter. 

Note that  one needs to know the  density profile of the dark matter in 
our galaxy to estimate the signal flux. 
A density profile of dark matter  may be parameterized as follows;
\begin{equation}
\rho(r)\propto\frac{1}{(r/a)^{\gamma}
\left[1+(r/a)^{\alpha}\right]^{(\beta-\gamma)/\alpha}}. 
\end{equation}
Here $r$ is the radial distance from the center of the 
galaxy and $\alpha, \beta$ and $\gamma$ are  free parameters 
to be fixed. 

The {\it modified isothermal distribution} is  parameterized as
$(\alpha,\beta,\gamma) = (2,3,0)$, which is 
consistent with visible star distributions near the 
center of the galaxy. 
Recently singular density 
profiles are proposed  and the implications are discussed. 
The singular distributions are motivated by  $N$ body 
numerical simulations and popular  parameterizations are 
those of 
Navarro Frenk White (1996)\cite{Navarro:1995iw} $(\alpha,\beta,\gamma) =(1,3,1)$
and 
 Moore (1999)\cite{Moore:1997sg} $(\alpha,\beta,\gamma) =(1.5, 3, 1.5)$. 
The simulations have been  significantly 
improved in recent years. The current best simulations 
contains typically $O(10^7)$ points, compared with 
$N=400$ in '70.  A better 'simulation' with large $N$ tends to predicts 
a more singular profile. 

The dark matter pair annihilation rate is proportional to 
$\rho^2$. A model with a  singular density profile at the center
of our galaxy predicts a  high dark matter 
pair annihilation rate. For example,
the  $\gamma$ ray  flux 
from the process $\tilde{\chi}\tilde{\chi}\rightarrow \gamma X$
is proportional to the 
$\rho^2$ integrated along the line of sight,
\begin{equation}
{\rm (signal\  rate)} \propto J(\psi)= 
\frac{1}{8.5{\rm kpc}}
\left(\frac{1}{\rm 0.3 GeV/cm^3}\right)^2
\int_{\rm line\  of\  sight}\rho^2(l) dl(\psi). 
\end{equation}
Note for  Moore's density profile, the integration of 
$\rho^2$ from $r=0$ is infinite, therefore all dark matter particles in the 
region $r<r_{cut}$ has been pair annihilated 
already\cite{Calcaneo-Roldan:2000yt}. To be quantitative 
\begin{equation}
\bar{J}(\Delta \Omega )\equiv \frac{1}{\Delta \Omega}
\int_{\Delta \Omega} J(\psi) d\Omega
\end{equation}
is 500 for the NFW profile and $10^5$ for  Moore's profile 
for a  typical angular resolution $\Delta\Omega= 10^{-3}$ sr. 
It was also  pointed out that the density profile not only 
has a cusp at the center, but 
it is also clumpy. The map of $J$ shows  numerous 
spots in the numerical simulations.

The $\gamma$ energy distribution  from the dark matter 
pair annihilation consisted of  two components. 
A ``continuum component''  comes from 
neutralino pair annihilation into $f\bar{f}$, $WW$, $ZZ$ and so on. 
The $\gamma$  energy distribution  is terminated at the kinematical limit 
$E_{\gamma}=m_{\tilde{\chi}}$. 
The other component comes from the neutralino pair annihilation 
into $\gamma Z $ and $\gamma \gamma$. Because a neutralino does not 
have a direct coupling to a photon, these processes are radiative, and 
suppressed. Energy of the photons is monochromatic 
$E_{\gamma} = m_{\tilde{\chi}}$.  
Diffuse gamma ray background is observed by EGRET\cite{Dixon:1998fk}.
Therefore the  continuum signal must be  observed as 
a kink structure  on the background spectrum. The monochromatic 
gamma signal is robust, because it cannot have any astrophysical 
origin.

EGRET covers $E_{\gamma}<30$~GeV. It has a photon angular  
resolution about 1.3 
degree at 1~GeV and 0.4 degree at 10~GeV. Sensitivity of the EGRET
data to the gamma ray flux from the center of the galaxy is 
$\sim 10^{-8}$ cm$^{-2}$ sec$^{-1}$ based on the operation time 
looking into the center of our galaxy.  
A future $\gamma$ ray observation
will be carried out by  GLAST\cite{Morselli:2002nw} 
(Gamma-ray Large Area Space Telescope) from 
2005. It has a 
sensitivity to the photon 0.02~GeV$<  E_{\gamma}<$ 300 GeV,
with the energy resolution $\Delta E_{\gamma} \sim$ 10\%. 
It  has a significantly better point source 
sensitivity compared with EGRET, 
$\Phi>10^{-10} {\rm cm}^{-2} s^{-1}$
for 1~GeV$< E< 300$~GeV.
``Ground Based'' ACTs (Airshower Cherenkov Telescopes) cover  
photons with O(10)~GeV$<E_{\gamma}<$ 10~TeV. Future
experiments such as VERITAS and  HESS
are planning  to have a sensitivity to the photon flux up to 
$\Phi= 10^{-13}{\rm cm}^{-2}s^{-1}$ for $E_{thr}>1$ TeV and
 $\Phi=5\times 10^{-14}{\rm cm}^{-2}
s^{-1}$ for 10 TeV.

Upper limit of the gamma ray flux from  the dark matter 
annihilation  comes 
from the continuum component. A typical dark matter pair annihilation 
produces O(10-40) low energy photons for $m_{\tilde{\chi}}=500$ GeV. 
The observed photon flux  from  the center of the galaxy 
set a limit on dark matter.    
EGRET has a sensitivity to the $\gamma$'s 
from the SUSY dark matter pair annihilation for 
$m_{\tilde{\chi}}<500$~GeV for  Moore's density  profile.
(See a recent improved analysis 
\cite{Hooper:2002fx} 
which takes into account the $E_{\gamma}$
dependence of the point source sensitivity.)

The gamma line  is a robust dark matter signal
if it is observed. The signal flux may be detectable 
for a  wino-like (expected for the anomaly mediated SUSY breaking model) 
or a Higgsino-like neutralino. Note that the pair annihilation cross
section into two $\gamma$s in the wino- or higgsino-limit is expressed as 
\begin{equation}
\sigma (\tilde{\chi}\tilde{\chi}
\rightarrow 2\gamma)\sim \frac{\alpha^2\alpha_2^2}{m_W^2}. 
\end{equation}
It does not have the usual $1/m^2_{\tilde{\chi}}$ dependence. 
This is because the mass of chargino in the loop satisfies  
$m_{\tilde{\chi}^+_1}\sim m_{\tilde{\chi}}$  for this case.
The large enhancement
also means that the perturbative calculations break down and 
all order resummations are required  to obtain the cross section. 
Recently  we calculated 
all order QED effect and 2 loop $W$ and $Z$ exchange effect 
\cite{Hisano:2002fk},
and the scale where perturbative approach breaks down is determined.
Summation of ladder exchange sometimes leads a 
huge enhancement of the cross section, and we are studying  all order 
corrections involving $W$ and $Z$ boson exchanges now. 

Anti-particle searches are also increasing their sensitivity
significantly.  Here the neutralino dark matter signal 
is anti-protons or positrons
produced by the pair annihilations. The annihilations occur
dominantly at the center of our galaxy, and the produced 
antiprotons propagate to our
solar system without too much loss.  Background secondary antiproton
flux is small in low momentum region.
Antiprotons coming from outside our solar system may be observed at
 the solar minimum. 

On the other hand, positrons loose their energy quickly by
synchrotron radiation and inverse Compton scattering in the 
Universe, therefore, 
they are  sensitive to  local clumps 
nearby our  solar system. Recently the HEAT
positron data  is interpreted as a dark matter signal and
implication is discussed by several groups\cite{Baltz:2001ir}.

The balloon-borne experiment BESS\cite{Orito:1999re} 
has been looking into  low energy
anti-protons in Canada. Duration of  each flight was typically a
few days and the data is statistically limited. The experiment is
going to move to
Antarctica, BESS-Polar.  They expect to fly in Jan 2004 and Feb 2006
with significantly improved flight time $\sim 20$ days. The second
flight is planned at the solar minimum.  The detector is sensitive down to
the antiproton with $ E_{\rm kin} \sim$ 100~MeV.  Another 
anti-matter search
will  be carried out by space-based AMS-02. This is an experiment at
International Space Station (ISS) and it was  planned to start from March
2004.\footnote{The 
experiment might be delayed due to the recent tragic space shuttle 
accident.} The threshold of  anti-proton 
kinetic energy is higher, $E_{\rm kin}> 300$~MeV, but 
long term operation for 3 to 5 years is possible. 
The statistics of positrons will be increased by a factor of 10 from 
HEAT experiment.

SUSY dark matter search is not the unique  target of these experiments. 
PBH or domain wall produce large antimatter signals.  Also
it is only very recently pointed out that solar modulation 
is charge dependent\cite{Bieber:1999dn}.  BESS confirmed the predicted 
quick increase of the   $\bar{p}/p $  ratio at the last solar maximum. 
In next few  years, the solar modulation is expected 
to be time dependent, and it is  interesting 
to  continue the observation of the $\bar{p}/p$ ratio.

\subsection{Solutions of the cusp problem}

There is a discussion that 
the number of the observed dwarf galaxies around the Milky Way halo is
not consistent with the $N$ body simulations. The discrepancy may be the
problem of numerical simulations. These simulations go through
complicated procedures such as interfacing a large scale simulation to
a small scale one. In addition, effects of radiations have not been
taken into account. However, the contradiction has been discussed from
various directions recently.

The cusp problem may be solved by introducing a warm dark matter 
with a free streaming scale $R_f\sim $ 0.1 Mpc. The 
warm dark matter washes out small scale density 
perturbations, so that cusps of the galaxy  do not have time to 
develop. For the gravitino dark matter, 
the required free streaming scale  corresponds to the gravitino 
mass around  $m_{3/2}\sim 1 $~keV, 
\begin{equation}
R_f=0.2(\Omega_W h^2)^{1/3}
\left(\frac{m_{3/2}}{\rm keV}\right)^{-4/3}
{\rm Mpc}. 
\end{equation}
On the other hand, 
the thermal relic density of O(1)~keV gravitino is too large,
\begin{equation}
\Omega_{3/2}h^2=0.5m_{3/2}({\rm keV}). 
\end{equation}
Therefore 
some entropy production to dilute the gravitino dark matter is 
required. A neutralino dark matter also could be a warm dark matter 
if it is produced from
topological defects \cite{Lin:2000qq}
or moduli(or heavy gravitino) decays. However, because 
the scattering of the neutralino with the  medium reduces 
the energy, one needs 10 TeV initial  neutralino energy at
very late time  $T_I<5$MeV\cite{Hisano:2000dz}.
However, we  need the density perturbation at the small scale
to have sufficiently early re-ionization of the universe. 
The constraint from Lyman-
$\alpha$  forest $z_{r.i.}\gsim 3 $  requires
$m_{3/2}>750$ eV for gravitino dark matter\cite{Narayanan:2000tp},
while the recent WMAP data\cite{Bennett:2003bz} 
push back the re-ionization period significantly $z_{r.i.}
=20^{+10}_{-9}$.
Another solution is found in \cite{Kamionkowski:1999vp}, 
where they propose an inflation model with suppressed density perturbation 
at small scale.

\section{On  the ground}
\subsection{$\Omega$ and MSSM parameter measurement}
We have seen 
many experimental and theoretical aspects related to 
SUSY dark matter searches in the previous section . 
On the other hand, the particle of which  dark matter cosnsists can be 
produced and studied at collider experiments.
Large Hadron collider 
(LHC) is a $pp$ collider at $\sqrt{s}=14$~TeV.   
Experiments are expected to start in 2007.
Significant SUSY parameter space would be 
covered within a year of operations.
In addition to that, 
$e^+e^-$ colliders at $\sqrt{s}= 500$~GeV to 
1~TeV are proposed 
by DESY, KEK and 
SLAC\cite{Aguilar-Saavedra:2001rg,Abe:2001gc,Abe:2001nn}. The LC 
will be a 
powerful tool to determine sparticle interactions.  

The thermal relic density of the neutralino dark matter  is  calculated 
by the following equation, 
\begin{equation}
\Omega_{\tilde{\chi}}^{th} h^2 \simeq 1.07\times 10^9 \frac{x_f  {\rm GeV}^{-1}}
{\sqrt{g_{*}} m_{pl} (a+ 3b/x_f)}
\end{equation}
Here $x_f=m/T_F$ and $T_{F}$ is the temperature where the neutralino 
decouples from the thermal equilibrium. The parameters 
$a$ and $b$ are related to 
the total lightest neutralino pair annihilation 
cross section at low energy which 
is expressed by the expansion in the relative velocity $v$, 
$\langle \sigma v\rangle\equiv a+ b v^2$.

$\Omega^{th}_{\tilde{\chi}}$  depends on the sparticle mass  spectrum. 
SUSY parameter space is strongly constrained if $\Omega_m \sim 
\Omega^{th}_{\tilde{\chi}}$ is assumed. 
The dark matter constraint is  studied 
extensively in the MSUGRA. 
The MSUGRA model is parameterized by scalar mass $m_0$, 
gaugino mass $M_0$, trilinear 
coupling $A_0$ and $\tan\beta$ at the GUT scale 
and SUSY mass spectrum at weak scale can be calculated from these 
parameters by solving the SUSY renormalization group  equations. 

The neutralino pair annihilation process  is  controlled by 
the following weak scale parameters at the low energy scale. 

\begin{enumerate}
 \item  The lightest neutralino  and slepton masses  
$m_{\tilde{\chi}}$ , $m_{\tilde{l}}$

In the minimal supergravity model, $\tilde{l}$ is lighter 
than $\tilde{q}$,  and the LSP is $\tilde{B}$-like unless 
$m_0\gg M_0$.  Therefore the neutralino pair annihilation into 
leptons through  $t$-channel exchange of a slepton is 
likely a dominant process.  
 $\Omega^{th}_{\tilde{\chi}} \propto m^4_{\tilde{l}}/m^2_{\tilde{\chi}}$.
       
\item The Higgsino component of the lightest neutralino. 

The amplitude  of the neutralino pair annihilation 
through $s$-channel exchange 
of a Higgs boson is proportional to 
$N_{\tilde{H}} N_{\tilde{B}(\tilde{W})}$. Also, $\sigma(\tilde{\chi} 
\tilde{\chi}\rightarrow WW)$ depends on the coupling to 
a $W$ boson which is proportional to  $N_{\tilde{H}}^2$. 
Therefore a  large higgsino component leads to  
small $\Omega^{th}_{\tilde{\chi}}$.

\item Higgs bosons with the mass close to 2$m_{\tilde{\chi}}$

If a neutralino pair annihilation hits the $s$-channel 
pole, the pair annihilation cross section becomes 
very large.  The amplitude 
${\cal M}$ is proportional to $ 1/(4m^2_{\tilde{\chi}}-m^2_P)$
in that case. 
This happens for large  $\tan\beta$ in MSUGRA .

\item Nature of sparticles with the masses close to $m_{\tilde{\chi}}$

If $m_{\tilde{\chi}} $ and the next lightest sparticle mass (NLSP) is 
degenerate, the co-annihilation of the LSP and the NLSP
cannot be neglected at the time of the decoupling. 
The relevant co-annihilation processes discussed in the literature 
are $\tilde{\chi}\tilde{\chi}^+_1$   
$\rightarrow VV'$, $\tilde{\chi}\tilde{\tau}\rightarrow \tau \gamma$
and so on. Such co-annihilation processes can be  O(100) faster 
than the pair annihilation process, and reduce the 
relic density significantly.
\end{enumerate}

The relevant masses and mixings need to  be constrained 
precisely to calculate the  thermal relic density of the LSP.
If this can be done, we can compare $\Omega^{th}_{\tilde{\chi}}$ with  
$\Omega_{m}$. They need not to be the same, 
because the calculation of the thermal relic density assumes 
the neutralino was once in thermal equilibrium 
and there was no entropy production after the decoupling. For example 
$\Omega_{m}<\Omega^{th}_{\tilde{\chi}}$ 
if there are late decaying particles.  
On the other hand, 
the neutralino  dark matter may be produced from the heavy particle 
(such as gravitino or 
moduli) decays after the neutralino decoupling\cite{Moroi:1999zb}, 
in that case, $\Omega_{ m}>\Omega^{th}_{\tilde{\chi}}$.  
Finally  the 
LSP dark matter may co-exist with other stable 
particles such as axion,  then 
$\Omega_{m}>\Omega^{th}_{\tilde{\chi}}$. 

The potential of the future collider experiments to determine 
SUSY parameters has been studied extensively. The main  motivation 
is to understand the origin of  supersymmetry breaking. 
The SUSY breaking in the MSSM sector must  originate from the SUSY 
breaking in  a ``hidden sector''.
Sparticle mass spectrum contains information on 
the SUSY breaking and mediation mechanism.  
The mechanism  may involve the gravitational interaction, 
 geometry of the higher 
dimensional space or  new interactions, namely the 
physics at much higher than TeV scale. 
Here we use  techniques to determine the 
SUSY breaking parameters in order to  estimate our knowledge 
about  $\Omega^{th}_{\tilde{\chi}}$ in future. 
 
In the collider experiments, the LSP is not directly visible because 
it is  neutral and stable.
The nature of the LSP would be studied by looking into 
production of heavier sparticles which 
decays into the LSP.

\subsection{LHC SUSY studies}
Squarks $\tilde{q}$ and gluinos $\tilde{g}$ are produced 
in the high energy $pp$ collisions at LHC with large cross sections.
They would further decay into jets( +leptons) + LSPs, and 
LSPs escape from detection. 
Isolation of events from a single 
cascade decay chain plays a key role for  sparticle mass 
determinations, because the 
decay distributions depend on the sparticle masses involved in 
the decay processes. 
For example, events with  high $p_T$ jets and opposite sign 
and same flavor leptons may be from the decay cascades 
$\tilde{q}\rightarrow j\tilde{\chi}^0_2\rightarrow\tilde{l} jl
\rightarrow \tilde{\chi}^0_1jll$. 
The $m_{ll}$ distribution 
is expressed as $d\Gamma/dm_{ll}\propto m_{ll}$ as can be seen in
Fig.~1. The end points of the $m_{ll}$ distribution is a
function of $m_{\tilde{\chi}}$, $m_{\tilde{\chi}^0_2}$,
and $m_{\tilde{l}}$ and can be measured with an error 
less than 1~GeV.  The measurement of 
the end point of the $m_{jll}$ and $m_{jl}$ distribution  
for a selected jet  and the leptons 
constrain $m_{\tilde{q}}$, $m_{\tilde{\chi}}$, $m_{\tilde{\chi}^0_2}$,
and $m_{\tilde{l}}$
\cite{Hinchliffe:1996iu}. 
The errors on slepton mass and neutralino mass 
are found to be  10\% in ATLAS TDR study\cite{AtlasTDR:1999fq}
with strongly correlated error. See Fig.~1(right).
This corresponds to $\Delta \langle\sigma v\rangle/\langle\sigma
v\rangle
\sim $ 20\%. 

LHC also has a sensitivity  to the higgsino and gaugino mixing of 
the LSP.  A squark can decay into the heaviest neutralino 
$\tilde{q}\rightarrow \tilde{\chi}^0_4$ followed by $\tilde{\chi}^0_4
\rightarrow\tilde{l}l \rightarrow \tilde{\chi}ll$. 
The end point of the 
$m_{ll}$ distribution may be  measured beyond the 
end point of $\tilde{\chi}^0_2\rightarrow l \tilde{l}\rightarrow 
ll \tilde{\chi}$ with 
O(5)~GeV error. If $M_1<M_2<\mu$ as in  MSUGRA, 
the end point gives us information on the $\mu$ parameter, 
and it strongly 
constrains the  higgsino component
of the lightest neutralino\cite{Drees:2000he}.

\begin{figure}[htbp]
\epsfxsize=5cm
\centerline{\epsfbox{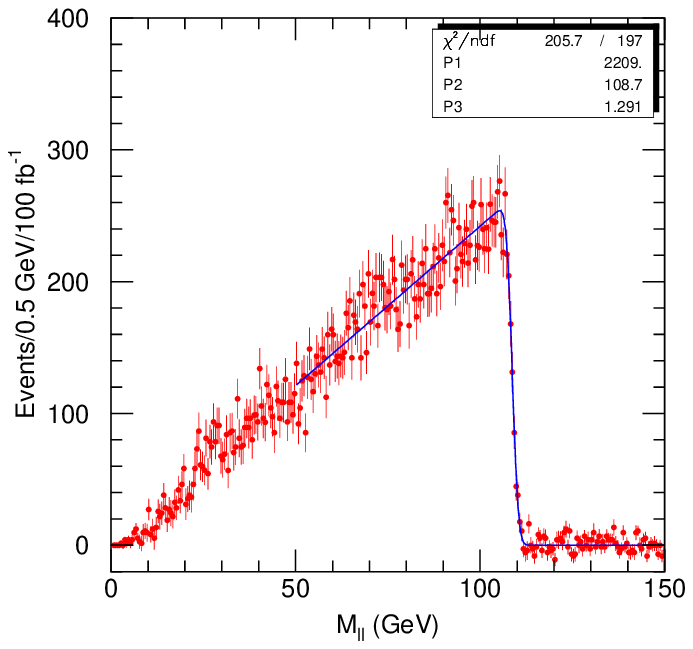} \hskip 1cm 
\epsfxsize=5cm
\epsfbox{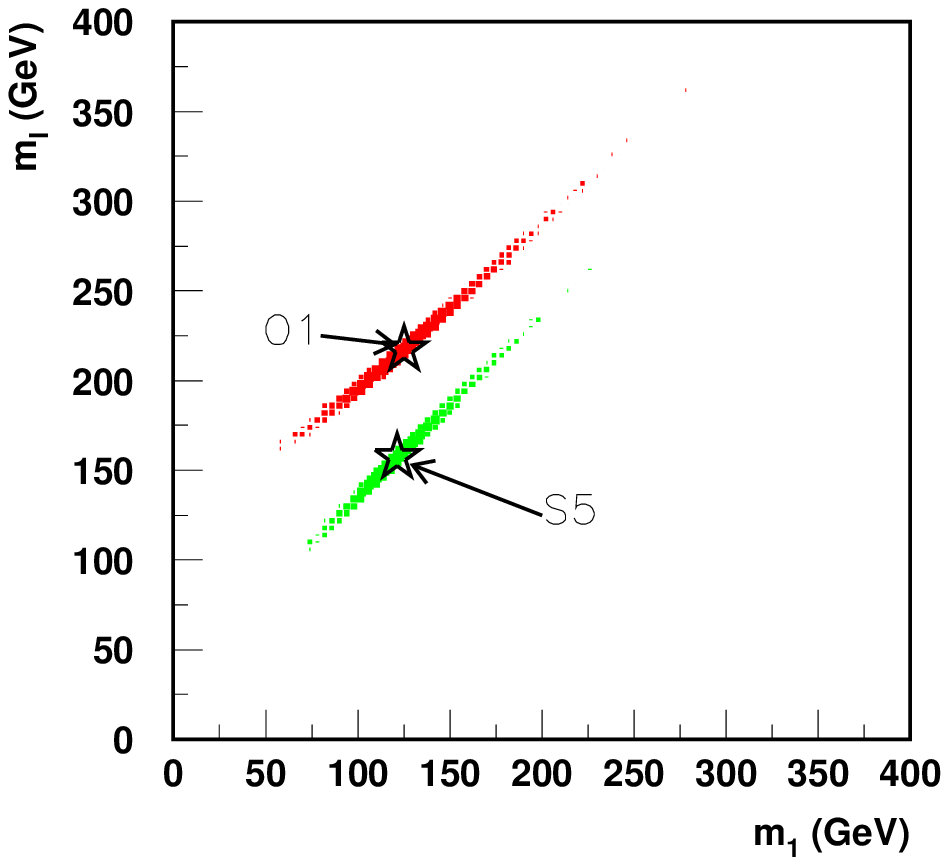}}
\caption{The $e^+e^- + \mu^+\mu^- -e^{\pm}
{\mu}^{\mp}$ mass distribution for LHC minimal supergravity Point 5 with
$\tilde{\chi}^0_2\to\tilde{l}l \to \tilde{\chi}^0_1
l^+l^-$~\protect\cite{AtlasTDR:1999fq}.  Right: Scatter plot of
reconstructed values of the $\tilde{l}_R$ and $\tilde{\chi}^0_1$
masses for LHC point 5 (S5) and for a different model (O1) using the
decay chain $\tilde{q}_L \to \tilde{\chi}^0_2 q \to \tilde{l}_R l q
\to \tilde{\chi}^0_1 llq$~\protect\cite{Allanach:2000kt}.}
\end{figure}

It is difficult to access all of the MSSM  parameters by the 
LHC alone. 
However, if we assume  MSUGRA, the parameters 
$ m_0, M_0, A_0$ and  $\tan\beta$ are precisely 
determined. The errors  could be around 1\% for $m_0$ and 
$M_0$ 
from slepton and ino mass measurements. 
$\tan\beta$ is determined  through 
the $\tilde{\tau}$ mass and Higgs sector. 
$A_0$ measurement is rather difficult, because it
does not affect the low energy spectrum except 
stop masses.  We recently showed that measurement of 
the decay distribution of  
$\tilde{g}\rightarrow (\tilde{t}t$  or $\tilde{b} b)$ 
$\rightarrow tb \tilde{\chi}^+_1$ constrain $A_0$ 
with reasonable 
accuracy\cite{Hisano:2003qu}. Within the MSUGRA assumption, the 
$\Omega^{th}_{\tilde{\chi}}$ may be determined within 
a few percent accuracy.

\subsection{Future $e^+e^-$ linear colliders}

Building a LC at $\sqrt{s}=500$~GeV 
to 1~TeV has been proposed and the physics at the LC
has been studied world-wide. Sparticles with  masses lighter than 
250~GeV are accessible at the first stage of the LC
experiments. The mass reach is significantly lower than that of LHC, where 
squark and gluino with $m_{\tilde{q}}, m_{\tilde{g}}<2$~TeV can be 
discovered.  However in the model with the universal gaugino mass 
at the GUT scale, such as MSUGRA,  charginos and neutralinos are significantly 
lighter than gluino, 
$M_1 : M_2 : M_3(\equiv m_{\tilde{g}}) = 0.4: 0.8 : 2.7$. 
Therefore the accessible parameter space for the LC at $\sqrt{s}\geq
1$ TeV is not significantly smaller  compared with that for  LHC. 

Weakly interacting sparticles are dominantly  produced at the LC. 
The background is less severe compared with LHC. 
Furthermore the dominant background such as $W^+W^-$ pair production 
can be  controlled  by a polarized electron beam.
Precise measurements 
of the masses and cross sections are possible for all kinematically 
accessible sparticles with typical errors of O(1)\% or less. 
The mass precisions through threshold scans are estimated typically as
\cite{Aguilar-Saavedra:2001rg}
\begin{eqnarray}
\Delta m_{\tilde{\chi}^{\pm}_1}, \Delta m_{\tilde{\chi}}&
\sim& 0.3 {\rm GeV},\cr
\Delta m_{\tilde{l}}\sim \Delta m_{\tilde{\nu}}&\sim& 0.1 {\rm GeV}, \cr
\Delta m_{\tilde{\tau}}&\sim& 0.6 {\rm GeV}.
\end{eqnarray}

\begin{figure}[htbp]
\epsfxsize=5cm
\centerline{\epsfbox{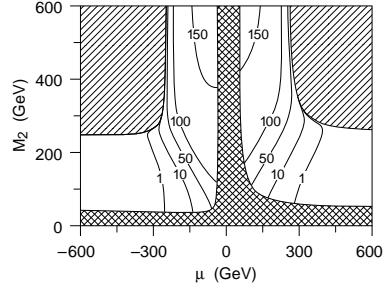}}
\caption{Contours of $\sigma(e^+e^-_R \to
\tilde{\chi}^+_1 \tilde{\chi}^-_1)$ (in fb) in the $(\mu, M_2)$ plane
for fixed $\tan\beta=4$ and $\sqrt{s}=500~{\rm GeV}$.  The cross-hatched
region is excluded by current bounds, and charginos are kinematically
inaccessible in the hatched region~\protect\cite{Feng:1995zd}. }
\label{feng}
\end{figure}

The production cross sections of the sparticles also depend strongly 
on the MSSM parameters.  
For example, even if  $M_1, M_2 \ll \mu$, and 
higgsino-like inos are not kinematically accessible at LC, 
the $\mu$ parameter 
would be constrained by measuring the 
the chargino pair
production cross sections for a polarized electron beam. 
In Fig.~\ref{feng} the $\sigma(e^+e^-_R\rightarrow \tilde{\chi}^+_1
\tilde{\chi}^-_1)$
is shown in $\mu$ and $M_2$ plane.
The cross section shows strong dependence on  $\mu/M_2$ ratio.
The clean environment at LC is also good to search 
for sparticles which are mass degenerate with the LSP. 
Finally combination of  LHC and LC data will improve the 
SUSY parameter study significantly\cite{LHCLC}.

\subsection{Gravitino and collider physics}
As it was mentioned already, 
a light gravitino with mass  $\sim$ O(1)~keV provides an 
interesting solution for the cusp problem. 
At collider experiments, such a  light gravitino 
cannot be produced directly, but it arises from 
the NLSP decay. When the lightest neutralino is 
the NLSP, the life time of the neutralino $c\tau$ 
is expressed by the gravitino mass $m_{3/2}$,
\begin{equation}
c\tau(\tilde{\chi}\rightarrow \psi_{3/2}) 
\sim 24m\left(\frac{100{\rm GeV}}{m_{\tilde{\chi}}}\right)^5 
\left(\frac{m_{3/2}}{\rm {1keV} }\right)^2. 
\end{equation}
The expression for the other NLSP sparticles are similar.

LHC is sensitive to the NLSP decays. 
If the NLSP is a charged particle, the measurements of the 
track of the charged NLSP  improve the 
SUSY parameter measurement. On the other hand, 
if the NLSP is neutralino, the decay into $\gamma$
would be  found  for  $c\tau< 1$km.
The error of the NLSP life time measurement, or equivalently  the error
of the gravitino mass at LHC is not yet fully explored. 
For LC $c\tau$ between 
$\rm O(10)\mu m <c\tau<O(10)m $ can be measurement within 10\%\cite{Ambrosanio:1999iu}.

\section {Outlook}
We discussed 
two directions to study  the lightest neutralino SUSY dark matter.

Because the SUSY dark matter has significant weak interactions, it can 
be searched for through  $\tilde{\chi}N$
scattering, or through the observation of exotic cosmic rays 
$\gamma, \bar{p}$ or $e^+$ arising from  neutralino pair annihilation 
in the galaxy. Although the search will cover significant 
MSSM parameter space in coming 10 years, there is large
uncertainty in the dark matter signal rate, because the dark matter 
density profile is not known. 
It is therefore difficult to interpret the observation, especially 
when searches appear to be negative.

On the other hand, the same dark matter particle will be produced at 
future colliders. 
The interaction of the 
lightest neutralino will be measured within  reasonable errors
if both LHC and LC are build. 
The  measurement would be useful to estimate the thermal 
relic density of the neutralino in the  
universe, and lower limits on the cross sections relevant to the 
neutralino dark matter searches. These measurements  will provide solid 
bases to understand the SUSY dark matter, 
the density profiles in our galaxy, and the 
thermal history of our universe.  The situation is 
sketched in Fig. 3. The science of the dark matter will
proceed through the interplay between particle physics 
and astrophysics, and it is a promising field even in the 21st century.

\begin{figure}
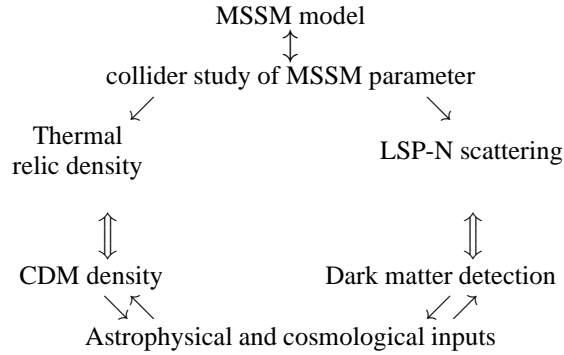

\begin{center}
MSSM model 
\\
{\large
$\updownarrow$}
\\
collider study  of MSSM parameter
\\
$\swarrow$ \hskip 3.5cm $\searrow$ 
\\
\begin{tabular}{c}
Thermal \cr
relic density
\end{tabular}
\hskip 3cm    
\begin{tabular}{c}
LSP-N scattering
\end{tabular}
\end{center}
\begin{center}
$\Big\Updownarrow$ \hskip 4.5cm $\Big\Updownarrow$ \\
\begin{tabular}{c}
CDM density
\end{tabular}
\hskip 2cm 
\begin{tabular}{l}
Dark matter detection 
\end{tabular}
\\
$\searrow$$\nwarrow$ \hskip 3.5cm 
$\swarrow$$\nearrow$ 
\\
Astrophysical\  and\  cosmological\  inputs
\end{center}
\caption{The road to understanding supersymmetric dark matter. From
\protect\cite{Feng:2002tc}. }
\end{figure}

{\bf Acknowledgment:}
I thank S.~Asai and N.~Sugiyama for their help when 
this lecture was first prepared two years ago, B.~Allanach and 
J.~L.~Feng for plots, M.~Drees for careful reading of the manuscript and  
 S.~Matsumoto, J.~Hisano and K.~Kawagoe for collaborations.
I also thank Prof. T.~Yamanaka for his kind encouragement 
on Dec. 31, 2002, when I started preparing this talk.
This work is supported in part by the Grant-in-Aid for 
Science Research, Ministry
of Education, Science and Culture, Japan 
(No.14540260 and 14046210).

\end{document}